\documentstyle[aps,floats,rotate,multicol]{revtex}

\input epsf         
\epsfverbosetrue    
\def\postscript#1{\begin{center}\leavevmode
\hbox{\epsfxsize=0.775\columnwidth\epsfbox{#1}}\end{center}}

\def\va{{\bf 1}}
\def\vb{{\bf 2}}

\def\vr{{\bf r}}

\def\vp{{\bf p}}
\def\vq{{\bf q}}

\def\vt{{\bf t}}
\def\vv{{\bf v}}

\def\vl{{\bf l}}
\def\vA{{\bf A}}
\def\vH{{\bf H}}

\begin{document}

\twocolumn[\hsize\textwidth\columnwidth\hsize\csname@twocolumnfalse%
\endcsname

\title{Superconducting zero temperature phase transition in two
dimensions and in the magnetic field}
\author{Wen-Chin Wu\cite{address} and Igor F. Herbut}
\address{Department of Physics, Simon Fraser University, Burnaby,
British Columbia, Canada V5A 1S6}

\date{\today}
\maketitle
\draft

\begin{abstract}
We derive the  Ginzburg-Landau-Wilson theory for the superconducting
phase transition in two dimensions and in the magnetic field.
Without disorder the theory describes a fluctuation induced
first-order quantum phase transition into the Abrikosov lattice. We
propose a phenomenological criterion for determining the transition field
and discuss the qualitative effects of disorder.
Comparison with recent experiments
on MoGe films is discussed.
\end{abstract}

]


\section{Introduction}
\label{sec:intro}

In the last decade it became well appreciated that the superconducting
transition can be tuned not
only by changing temperature, but also by varying some other parameter,
like the level of disorder, magnetic field, film thickness, or doping of
a high-$T_c$ material, at {\it zero temperature} ($T=0$) \cite{goldman}.
This is an example of a {\it quantum} phase transition
\cite{subir}, where the order-disorder transition is not brought about
by increasing the entropy of the system, as in the finite temperature case,
but by changing the quantum mechanical ground state in a fundamental way.
A particularly
interesting example is the superconductor-insulator transition in two
dimensions (2D) \cite{herbut2000}, where the superconducting state
at $T=0$ is destroyed by increasing the level of static disorder in
the system. For a while it has been thought that the magnetic field
would have a similar (but not exactly the same)
disordering effect on the superconducting ground state
\cite{goldman,fisher,yazdani};
recently, however, the experiments of Mason and Kapitulnik \cite{mason99}
have suggested that
by increasing the magnetic field at $T=0$ the 2D system
undergoes a superconductor-metal phase transition, and,
moreover, that this transition may actually be discontinuous.
The metallic state they found
seems to be quite unusual, in that its resistivity
is anomalously low, and it exhibits a very large magnetoresistance.
Only at a higher magnetic field the expected 2D insulating state is
recovered \cite{mason99}.

Motivated by these intriguing results, in this paper we study
theoretically the $T=0$ superconducting phase transition in 2D and in
the perpendicular magnetic field. We begin by demonstrating that, due to the
pair-breaking nature of the magnetic field, in 2D (and
without disorder) there exists a regular
Ginzburg-Landau-Wilson (GLW) theory for the fluctuating
superconducting order parameter
even at $T=0$. Long time ago, it was shown by
Maki and Tsuzuki \cite{maki65} that the GLW theory in the
magnetic field and in 3D is almost regular at $T=0$; the
quartic term coefficient was singular only weakly (logarithmically) as
$T$ approaches zero. We show that in 2D this
singularity is removed by the absence of the third
direction along which the motion would be unaffected by the field.
The resulting GLW theory describes the
quantum fluctuations of the lowest Landau level (LLL) superconducting
order parameter (OP), and bears close resemblance to the finite-$T$ GLW
theory for a 3D superconductor. Based on the existing understanding
of the finite-$T$ Abrikosov transition in 3D, we then argue that the quantum
Abrikosov transition in 2D and without disorder
is generically first-order, and propose a phenomenological
criterion for determining the first-order superconductor-metal
transition field $H_{sm}$. In particular, we show
that the true transition field $H_{sm}$ at $T=0$ is always below
the mean-field critical field $H_0$, and that the quantum
phase transition is from the superconducting
into a zero-temperature equivalent of the vortex liquid phase:
a metallic phase, strongly renormalized by the superconducting
fluctuations, which we propose may be related to the
anomalous metal observed by Mason and Kapitulnik.

In a  real experiment, of course, disorder is always present, and may actually
be quite strong, so it can be expected to play an important role in
determining the nature of the transition. In principle,
quenched disorder lifts the degeneracy of the LLL for the
order parameter, and, we argue, if strong enough
could bring back the
continuous phase transition to a glassy superconducting
phase. We provide a qualitative discussion of this phase transition
at strong disorder within the framework of a
self-consistent Hartree approach. The superconducting transition
in this approach would correspond
to the condensation into the lowest extended Hartree eigenstate, that
in principle needs
to be computed numerically. At weak disorder, on the other hand, based
on the existing renormalization group studies of the thermal Abrikosov
transition, we expect that the superconducting transition at $T=0$ remains
discontinuous.

The organization of the paper is as follows. In Sec.~\ref{sec:action},
we derive
the effective GLW action appropriate for the superconductor-metal
quantum phase transition in 2D in the
magnetic field and without disorder.
In Sec.~\ref{sec:transition}, we show how this effective
action leads to a fluctuation induced first-order transition, and calculate
the critical field.
In Sec.~\ref{sec:disorder}, we discuss the qualitative effects of
quenched disorder
on the phase transition, and in Sec.~\ref{sec:conclusion}, we relate
our results to experiment and other theoretical studies
and provide a brief summary.
Technical details are relegated to the Appendices.

\section{Quantum action at $T=0$}
\label{sec:action}

\begin{figure}[t]
\postscript{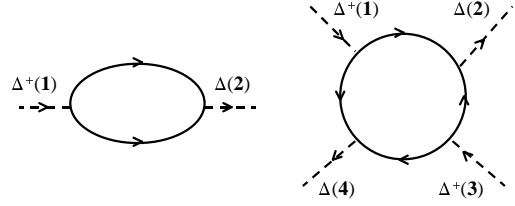}
\caption{Diagrammatic representation of the quadratic $S_2$ and the
quartic $S_4$ terms of the action in (\protect\ref{eq:action2}). Solid
lines correspond to single-particle Green's functions.}
\label{fig1}
\end{figure}

We begin by considering the $T=0$ action
for a system of 2D fermions in a magnetic field
($\hbar=c\equiv 1$)

\begin{eqnarray}\label{eq:action}
S&=&\sum_\sigma\int d^2\vr d\tau\left[\psi^\dagger_\sigma\partial_\tau\psi_\sigma
+{1\over 2m}|(\nabla-ie\vA)\psi_\sigma|^2-\mu\psi^\dagger_\sigma\psi_\sigma\right.
\nonumber\\
&-&\left.{V\over 2}\psi^\dagger_\sigma\psi^\dagger_{-\sigma}\psi_{-\sigma}\psi_\sigma
\right],
\end{eqnarray}
where $\psi\equiv \psi(\vr,\tau)$, $\sigma$, $e$, and $m$ denote the
spin, charge, and
mass of fermions, $\mu$ is the chemical potential, and
$\vH=\nabla\times\vA$ is the external magnetic field. The integration
over the imaginary time $\tau$ is over the whole real axis ($T=0$),
and we neglect the Zeeman coupling of the magnetic field to spin.
We assume the interaction $V>0$ which corresponds to the $s$-wave attraction.
The partition function is the functional integral over Grassman fields $\psi$
with the action $S$ as the Boltzmann weight.
Following the standard Hubbard-Stratonovich
decoupling procedure \cite{hertz76} in the Cooper channel,
the action can be rewritten as

\begin{eqnarray}\label{eq:action1}
S&=&\int d^2\vr d\tau\left[\psi^\dagger_\sigma\partial_\tau\psi_\sigma
+{1\over 2m}|(\nabla-ie\vA)\psi_\sigma|^2-\mu\psi^\dagger_\sigma\psi_\sigma
\right.\nonumber\\
&-&\left.|\Delta(\vr,\tau)|^2-\sqrt{V}\Delta(\vr,\tau)
\psi^\dagger_\uparrow\psi^\dagger_{\downarrow}-\sqrt{V}\Delta^\dagger(\vr,\tau)
\psi_{\downarrow}\psi_\uparrow\right]
\end{eqnarray}
in terms of the fluctuating superconducting
order parameter (OP) $\Delta(\vr,\tau)$.
(The sum over the repeated spin indices is here assumed.)
Since the action now became quadratic in the fermionic
fields, in principle they could be integrated out so that one is left with
the action in terms of the OP only. Furthermore, one could attempt to
expand the resulting action in powers of the OP,

\begin{eqnarray}\label{eq:action2}
S&=&S_2+S_4+\cdot\cdot\cdot~ =\int d{\va} d{\vb}~ K_2({\va},{\vb})
\Delta^\dagger({\va})\Delta({\vb})\nonumber\\
&+& \int d{\bf 1}d{\bf 2}d{\bf 3}d{\bf 4}~ K_4({\bf 1},{\bf 2},{\bf 3},{\bf 4})
\Delta^\dagger({\bf 1})\Delta({\bf 2})\Delta^\dagger({\bf 3})\Delta({\bf 4})\nonumber\\
&+&\cdot\cdot\cdot,
\end{eqnarray}
where we denoted ${\bf 1}\equiv (\vr_1,\tau_1)$ for brevity.
The partition function is the functional integral over all
configurations of the {\it fluctuating} OP.
The quadratic and the quartic terms in the action can be
represented diagrammatically as in Fig.~\ref{fig1}. As well known,
without the magnetic field
these diagrams would lead to singular expressions for $K_2$ and $K_4$ at
$T=0$, as  the system would always be unstable towards the
formation of Cooper pairs. Finite magnetic field, however, acts as a
pair-breaker, and one may expect it to
regularize the kernels in the GLW expansion, similarly
as disorder does for the $d$-wave pairing \cite{dwave}. In the rest
of this section we will show how this happens in 2D, and obtain the
GLW theory for the superconducting transition at $T=0$.

In the  magnetic field the momentum is no longer a good
quantum number, but the gauge invariance has to be maintained in the
theory. This dictates that instead of the standard
gradient expansion, the kernels in the GLW
action can now be expanded in powers
of the covariant derivative $\nabla-ie^*\vA$, with $e^*=2e$.
Furthermore, assuming a weak coupling $V$,  the transition into the
superconducting state can be expected at a low magnetic field, so to
calculate the diagrams in Fig.~\ref{fig1} one may follow Gor'kov and use
the ``semi-classical'' approximation \cite{gorkov,agd},  for
the single-particle Green's function

\begin{equation}\label{eq:semiclass}
G_\nu(\vr_1,\vr_2)=e^{i\phi(\vr_1,\vr_2)}G^0_\nu(\vr_1,\vr_2).
\end{equation}
Here $\nu$ is the Matsubara frequency. The phase is determined by
$\phi(\vr_1,\vr_2)=e\int_{\vr_2}^{\vr_1} \vA \cdot d\vl$,
integrating the gauge
field over the straight path from $\vr_2$ to $\vr_1$,
and $G^0$ is the propagator in the absence of magnetic field.
Finally, since the OP feels the magnetic field, one may expand it in terms
of the Landau levels, which are the solutions of the single-particle
Schr\"{o}dinger equation in the magnetic field for charge $2e$.
Near the superconducting transition
only the lowest Landau level (LLL) configurations
of the OP need to be retained
in the effective action \cite{abrikosov57}, since the higher Landau
levels are not critical, and their effect in principle could be
only to renormalize the
coefficients in the GLW expansion for the LLL modes.
Choosing the Landau gauge $\vA=H(0,x,0)$ to represent the
magnetic field $\vH=H\hat{z}$ perpendicular to the 2D plane,
the OP may then be expanded in terms of the LLL functions

\begin{equation}\label{eq:lll}
\Delta(\vr,\tau)\equiv\sum_{-\infty}^\infty C_n(\tau) e^{ikny}\psi_n(x);
~~~\psi_n(x)=e^{-eH(x-{kn\over
2eH})^2},
\end{equation}
where $k=2\pi/L$ ($L$ is the linear size of the system)
and $n$ is an arbitrary integer.

Going to Matsubara frequency space, the quadratic part of the action
can be written as
$S_2=\int d\omega~ \bar{S}_2(\omega)$, and to the lowest order
in small parameter $\omega\tau_H$ we find (see Appendix~\ref{appA})

\begin{equation}\label{eq:k2}
\bar{S}_2(\omega)=[1-g\ln(2\Omega\tau_H)+\sqrt{\pi}g\tau_H|\omega|]
\langle|\Delta_\omega(\vr)|^2\rangle,
\end{equation}
where $\tau_H=\ell/v_{\rm F}$ is the cyclotron time
with $\ell=1/\sqrt{2eH}$ the magnetic length for charge $2e$
and $v_{\rm F}$ the Fermi velocity.
$g= {\cal{N}}V/2$ is the dimensionless coupling
(${\cal{N}}=m/2\pi$ is the constant density in 2D),
$\Delta_\omega(\vr)$ is the Fourier transform of
$\Delta(\vr,\tau)$,
and $\Omega$ is the usual ultraviolet cutoff. Also

\begin{eqnarray}\label{eq:delta2}
\langle|\Delta_\omega(\vr)|^2\rangle&=&\int~d\vr|\Delta_\omega(\vr)|^2\nonumber\\
&=& {2\pi\over k}
\sum_n|C_n(\omega)|^2\sqrt{\pi\over 2eH}.
\end{eqnarray}
Similarly, the quartic part of the action
$S_4=\int d\omega d\omega_1 d\omega_2 \bar{S}_4(\omega,\omega_1,\omega_2)$,
where at zero frequencies \cite{approx} (see Appendix~\ref{appB}),

\begin{equation}\label{eq:k4}
\bar{S}_4(\omega=\omega_1=\omega_2=0)=a{g^2\tau_H^2\over {\cal{N}}}
\langle|\Delta_0(\vr)|^4\rangle
\end{equation}
with the constant $a=4\pi[\ln(1+\sqrt{2})]^2$ and the average

\begin{eqnarray}\label{eq:delta4}
&&\langle|\Delta_0(\vr)|^4\rangle=\int d\vr~
|\Delta_{\omega=0}(\vr)|^4\nonumber\\
&&={2\pi\over k}
\sum_{n,m,p,q}\delta(n-m+p-q)C_{n}(0)C^*_{m}(0)C_{p}(0)C^*_{q}(0)\nonumber\\
&&\times
\sqrt{\pi\over 4eH}e^{-{k^2\over 4eH}[(n^2+m^2+p^2+q^2)-(n+m+p+q)^2/4]}.
\end{eqnarray}

Note that the coefficient
before $\langle|\Delta_0(\vr)|^4\rangle$ is finite in a finite magnetic field.
At zero field it diverges, as well-known. In 3D the same coefficient is
$\propto -\ln (T/H)$, and
logarithmically divergent as $T\rightarrow 0$ \cite{maki65}.
This is a consequence of the fact that the motion in the $z$ direction
remains unaffected by the magnetic field. In 2D, however, the singular
behavior of the Landau expansion around the normal state $\Delta=0$ for the
$T=0$ action is completely cured.

The effective GLW action for the superconductor-metal phase transition in 2D
at $T=0$ can thus be written as

\begin{eqnarray}\label{eq:glw}
S_{\rm GLW}[\Phi]=\int d\vr d\tau \left[\Phi^\dagger|\partial_\tau|\Phi
+\alpha |\Phi|^2+\beta|\Phi|^4\right]
\end{eqnarray}
where $\Phi\equiv\Phi(\vr,\tau)$, and the partition function is
$Z=\int D[\Phi^*, \Phi] \exp(- S_{\rm GLW}[\Phi])$. In principle the action
contains higher powers of the field and the higher order time derivatives,
but we make the usual approximation in retaining only the most
relevant terms.
The operator $|\partial_\tau|$ corresponds to $|\omega|$ in Matsubara space
and we have rescaled the field
$\Phi=(\sqrt{\pi}g\tau_H)^{1/2}\Delta$
to bring the coefficient of the $|\omega|$ term in (\ref{eq:glw}) to unity.
Thus $\alpha=[g^{-1}-\ln(2\Omega\tau_H)]/(\sqrt{\pi}\tau_H)$ and
$\beta=a/(\pi {\cal{N}})$.

\section{Fluctuation induced first-order transition}
\label{sec:transition}

When the fluctuations of the OP are ignored (the mean-field approximation),
the system described by the action (\ref{eq:glw}) at $T=0$
undergoes a continuous phase transition from normal to the Abrikosov
superconducting phase as the magnetic field is decreased, at the point where
$\alpha=0$. The mean-field critical field is thus
$H_{0} =2\Omega^2/(ev_{\rm F}^2) e^{-2/g}$, and
exponentially small for a weak coupling $g$.
One can easily verify that this satisfies the standard relation $H_0
\xi^2 \approx 1$ where $\xi$ is the superconducting coherence length.
This mean-field critical field is simply the end ($T=0$) point of the
standard $H_{c2}(T)$
Abrikosov line in the $H$-$T$ phase diagram of a 2D
type-II superconductor.
However, today it is well established both theoretically
\cite{brezin85,tesanovic91,herbut94} and experimentally \cite{safar,hardy}, that
at finite $T$ Abrikosov transition is one of those rare cases where
fluctuations even {\it qualitatively} alter the nature of the transition,
turning it into a first-order in the clean case.
This may be understood as a consequence
of the macroscopic degeneracy of the LLL manifold, which is a unique
feature of the Abrikosov transition, and which greatly enhances the effect
of fluctuations. Any OP configuration within the LLL can be parametrized
in terms of its zeroes (vortices), and the Abrikosov transition is
equivalent to a freezing transition of a complicated classical many body
system, which is then typically first order \cite{herbut94}.
Our quantum action at $T=0$ is almost identical to the
finite-$T$ GLW theory for a 3D superconductor in the magnetic field
(except for a linear instead of a quadratic derivative
with respect to the third coordinate),
and one may expect that, in this case, quantum fluctuations of the
OP may also turn the
$T=0$ transition into first order. To approximately obtain the
critical magnetic field $H_{sm}$ of the first-order transition, it
is convenient to work with dimensionless quantities and
first rescale the fields, lengths, and time
as $(4\pi\ell^2\tau_H\beta)^{1\over 4}\Phi\rightarrow \Phi$,
$r/(\sqrt{\pi}\ell)\rightarrow r$, and $\tau/\tau_H\rightarrow \tau$.
The action in (\ref{eq:glw}) then becomes

\begin{eqnarray}\label{eq:glwr}
S_{\rm GLW}[\Phi]=\int d\vr d\tau \left[g_\tau\Phi^\dagger|\partial_\tau|\Phi
+g_\alpha |\Phi|^2+{1\over 4}|\Phi|^4\right],
\end{eqnarray}
where the two dimensionless couplings are $g_{\tau,\alpha}=(1/\tau_H,\alpha)
(\pi\ell^2\tau_H/4\beta)^{1/2}$.
First, consider only the OP configurations
that have no $\tau$ dependence ($\omega =0$ modes). For those
 the first term of
(\ref{eq:glwr}) vanishes, and the thermodynamics of the system would
depend exclusively on the single dimensionless coupling $g_\alpha$.
The theory would then look zero-dimensional, however, this is deceiving:
the macroscopic degeneracy of
the LLL still remains. In fact, without the first term the partition
function with the action
(\ref{eq:glwr}) would be identical
to the finite-$T$ partition function that describes the  vortex-liquid-to-solid
transition, and which is known to have a weak first-order transition
at $g_\alpha=g_{\rm M}\approx -6.7$. This has been established in
detailed Monte Carlo simulations \cite{tesanovic91}, as well as in
the density-functional theory \cite{herbut94}.
To find the transition field with the full $\tau$-dependence of the
OP included we assume that the transition
is driven by the same mechanism of growing positional
correlations between vortices, the only difference now being
that the vortices are ``straight'' in $\tau$-direction over an imaginary
time scale $\bar{\tau}$, instead of $\tau_H$.
We identify this time scale with the correlation ``length''
 in $\tau$ direction.
This ansatz works remarkably well in describing the finite-$T$
first-order transition line in clean YBCO \cite{herbut95},
and on that basis we expect it
to be a good approximation here as well.
The transition field is then determined by the condition

\begin{equation}\label{eq:melting}
g_{\rm M}=g_\alpha \bar{\tau}^{1\over 2},
\end{equation}
for $\bar{\tau}>1$ in units of $\tau_H$.
To approximately determine $\bar{\tau}$ we will use the self-consistent
Hartree approximation \cite{herbut95} to the action
in (\ref{eq:glwr}). The correlation time is then defined as

\begin{equation}\label{eq:tau}
\bar{\tau}= {g_\tau\over g_\alpha+\langle|\Phi(\vr)|^2\rangle/4},
\end{equation}
where the quantum mechanical
average appearing in (\ref{eq:tau}) is determined self-consistently as

\begin{equation}\label{eq:hartree}
\langle|\Phi(\vr)|^2\rangle={1\over 2\pi}\int  {d\omega\over g_\tau|\omega|+
g_\alpha+\langle|\Phi(\vr)|^2\rangle/4}.
\end{equation}
Solving Eqs.~(\ref{eq:melting})-(\ref{eq:hartree}), one obtains the equation
for the first-order transition field

\begin{equation}\label{eq:hc}
H_{sm} ={2\Omega^2\over ev_{\rm F}^2} e^{-2/g_{\rm eff}},
\end{equation}
where the effective coupling satisfies

\begin{equation}\label{eq:gbar}
g_{\rm eff}^{-1}=g^{-1}+{4a|g_{\rm M}|\sqrt{\ln\Omega_1}\over \pi}
{\sqrt{2eH_{sm}}\over k_{\rm F}},
\end{equation}
in the $\Omega_1\rightarrow \infty$ limit.
The parameter $\Omega_1$ is a dimensionless upper cutoff we
introduced to regularize the integral in (14). It
may be chosen to yield the correct answer
when the next order in ``frequency'' ({\em i.e.},
$\omega^2$) term is kept in the effective action, for example.
Since $\Omega_1$ appears only under a log the result is
little sensitive to its precise value.
More importantly, it is evident that $g_{\rm eff}$ is smaller than $g$
which in turn implies that the true transition
field $H_{sm}$ is below the mean-field $H_0$, which would correspond
to the $g_{\rm M}=0$ case.

\begin{figure}[t]
\postscript{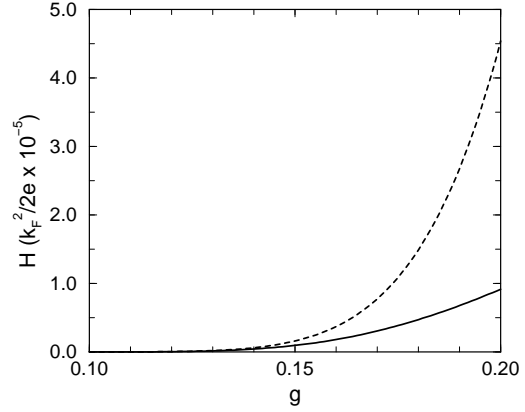}
\caption{$H$-$g$ phase diagram of superconductor-to-metal QPT
in 2D in a magnetic field. The dashed line corresponds to
the the mean-field second-order transition. The solid line is determined
by solving Eqs.~(\protect\ref{eq:hc}) and (\protect\ref{eq:gbar})
and corresponds to the first-order transition induced by
the quantum fluctuations of the superconducting order parameter.}
\label{fig2}
\end{figure}

In Fig.~\ref{fig2} we present the  $H$-$g$ phase diagram for
the 2D system in the  magnetic field and at $T=0$.
The superconductor-metal phase
transition induced by the order parameter fluctuations
is first-order. The mean-field result represented by
the dashed line is included for comparison.
Her we have assumed $\Omega=\epsilon_{\rm F}$, $\Omega_1=10^{4}$, and
$g_{\rm M}=-7.0$. The mean-field and the true transition field are
indistiguishably close at very small coupling, but they start to
differ significantly
at larger couplings. In principle, one may also expect a somewhat
renormalized value of $g_M$ from the static value ($\approx -7$),
but the difference does not qualitatively alter our results.

\section{Effects of disorder}
\label{sec:disorder}

So far  we have completely neglected the effects of static disorder on the
superconducting transition. Since the inclusion of disorder
complicates the problem significantly, we will discuss
its  effects only qualitatively. First, it should
be possible to model disorder by
including a random potential term $V(\vr)|\Phi(\vr)|^2$ into the OP
effective action in (\ref{eq:glwr}). That such a term indeed arises in the
OP theory was demonstrated by one of us \cite{dwave} in a related problem
of disordered $d$-wave superconductor. For simplicity one may assume that
the random potential is uncorrelated in space. With this term included
one faces
an interacting theory with quenched disorder (and in the LLL)
which is notoriously difficult to analyze. So in what follows we assume
that disorder is strong (compared to the quartic term), so that the
self-consistent Hartree treatment of the interaction may be a reasonable
starting point \cite{book}. In this spirit we replace the quartic term in (11) as

\begin{equation}
\Phi ^4 \rightarrow \langle \Phi ^2 \rangle \Phi^2,
\end{equation}
where the average is self-consistently determined as

\begin{equation}\label{eq:hartreei}
\langle|\Phi(\vr)|^2\rangle={1\over 2\pi}\sum_n
\int d\omega{|\phi_n(\vr)|^2\over
g_\tau|\omega|+\varepsilon_n},
\end{equation}
where $\phi_n$ are the LLL eigenfunctions, and $\varepsilon_n$
eigenvalues of the random potential, Hartree "screened" by the interactions

\begin{equation}\label{eq:seq}
\tilde{V}(\vr) = g_\alpha+V(\vr)+
{1\over 4}\langle|\Phi(\vr)|^2\rangle.
\end{equation}
The superconducting phase transition in this approach
would correspond to the vanishing of the
lowest eigenvalue $\varepsilon_n$ and condensation into the
corresponding extended random eigenstate \cite{tu}.  This transition is
expected to be continuous, but to verify this scenario one needs
to implement the self-consistent procedure numerically, which
we leave for future work. The reader should note however, that the
main effect of disorder is to lift the degeneracy of the LLL manifold,
and thus restore the possibility of a continuous transition.

In case of weak disorder the above Hartree approximation
becomes inadequate and we may
only speculate what happens with the transition. Our prejudice, which
we discuss more shortly,  is that
the transition remains first order for weak disorder, and turns
continuous only at some critical disorder strength.

\section{Summary and discussion}
\label{sec:conclusion}

Our conclusion of a first order transition in the clean case is
in agreement with the renormalization group studies of
Br\'{e}zin, Nelson, and Thiaville \cite{brezin85} and more recently
of Moore and Newman \cite{moore95}. These authors have shown that due to
macroscopic degeneracy of the LLL the renormalization group
flow in dimensions less than six is always unstable, which is
usually interpreted as a sign of a first-order transition. Although
they studied a thermal, and not quantum,
Abrikosov transition, their methods and conclusions can be readily
translated to our case. Furthermore, our conclusion is also in
agreement with the large-$N$ treatment of the Abrikosov transition
\cite{affleck,radz,herbut96,lopatin99}
at finite temperature. As for the disorder case, Moore and Newman
\cite{moore95} also demonstrated
that with weak disorder the renormalization group
flow remains unstable, which would
indicate that the transition is still discontinuous. The case of strong
disorder is outside the domain of validity of their perturbative approach,
and we feel that the self-consistent Hartree treatment of the
interaction term we discussed is more appropriate.

Our theory offers a natural explanation for the hysteretic behavior
observed at the $T=0$ superconductor-metal transition by Mason and Kapitulnik.
We argue that it is a consequence of the fluctuation induced first-order
transition. Also, that the non-superconducting state appears
metallic, and not insulating, may be related to the fact
that a weak magnetic field cuts off the weak-localization effects,
and thus relegates the localization effects in 2D to lower tempereatures
\cite{lee}. The observed
anomalously small resistivity and the large magnetoresistance
of the  Mason and Kapitulnik metal could
be related to the strong superconducting fluctuations near the
critical field.  Additional support for this picture comes
from the experimental observation that the resistivity away from the critical
field $H_{sm}$ vanishes as $R\sim (H-H_0)$, where $H_0 > H_{sm}$
\cite{mason99}.
This power-law follows naturally from our theory as follows.
Away from the critical field (outside the critical region
dominated by the OP fluctuations) the transition appears continuous, and
the resistivity should vanish according to the simple
(Aslamazov-Larkin like) scaling \cite{huse}

\begin{equation}
R\sim \xi ^{d-2},
\end{equation}
where $\xi$ is the superconducting correlation length. The
GLW action (10) for the OP at the mean-field level
implies $d=0$ (since the OP is confined to the LLL which completely
eliminates the gradient terms in 2D). At the mean-field level
$\xi \sim 1/(H-H_0)^{1/2}$, so the power counting implies

\begin{equation}
R\sim (H-H_0),
\end{equation}
in the crossover region,
precisely as seen in the experiment \cite{mason99}. This, of course,
ceases to be valid upon entering the critical region, in which,
as we argued, fluctuations eventually drive the transition first order
at a lower critical field $H_{sm}$.

Finally, we note that the GLW theory in Eq.~(10) has the frequency
dependence characteristic of a dissipative system. This is analogous
to what was found in the theory of disordered $d$-wave superconducting
phase transition at $T=0$ \cite{dwave},
and is related to the pair-breaking nature
of the magnetic field. This may provide the theoretical basis for
the Mason and Kapitulnik interpretation of the superconductor-metal
transition \cite{mason99}.

To summarize, we have studied the $T=0$  superconducting
phase transition for a 2D system in a magnetic field.
We derived an effective Ginzburg-Landau-Wilson action for the
fluctuating superconducting order parameter which
enables one to investigate the effects of quantum fluctuations
and the order of the phase transition. It is argued that
without, or with weak disorder, the quantum superconductor-metal
phase transition is of first order.
For the case of strong disorder we expect that the possibility of a
continuous phase transition is restored.

\acknowledgements
We are grateful to Prof. S. Sachdev and Dr. D. Lee
for useful discussions.
This work is supported by NSERC of Canada and NSC of Taiwan (under the grant No.
89-2112-M-003-027). IFH is also supported by an award from Research Corporation.

\appendix

\section{Derivation of $S_2$}
\label{appA}

In this Appendix
we give the explicit derivation of quadratic $S_2$ in (\ref{eq:k2}).
If we write $S_2=\int d\omega~ \bar{S}_2(\omega)$,
based on the diagram in Fig.~\ref{fig1},
one has at $T=0$ (apart from the constant term)

\begin{eqnarray} \label{eq:S2a}
&&\bar{S}_2(\omega)=-{V\over 2}\times\nonumber\\
&&\int{d\nu\over 2\pi} d\vr_1 d\vr_2 G_{\nu+\omega}(\vr_1,\vr_2)
G_{-\nu}(\vr_1,\vr_2)\Delta^\dagger_\omega(\vr_1)\Delta_\omega(\vr_2),
\end{eqnarray}
where $\Delta_\omega$ is given by (\ref{eq:lll}) with
$C_n(\tau)\rightarrow C_n(\omega)$  in terms of LLL.
In the semi-classical approximation,
the 2D single-particle Green's function $G$ is given by (\ref{eq:semiclass})
in which

\begin{eqnarray} \label{eq:G0a}
G_\nu^0(\vr_1,\vr_2)=\int{d\vp\over (2\pi)^2}e^{i\vp\cdot(\vr_1-\vr_2)}
{1\over i\nu-\epsilon_\vp}
\end{eqnarray}
where $\epsilon_\vp=\vp^2/2m-\epsilon_{\rm F}$ and
the phase $\phi(\vr_1,\vr_2)=eH(x_1+x_2)(y_1-y_2)/2$. We have
assumed that the field $\vA=H(0,x,0)$, which leads to $\vH=H\hat{z}$.
After substitution of the above into (\ref{eq:S2a}) and carrying
out the integrations over $y_1$ and $y_2$, one obtains

\begin{eqnarray} \label{eq:S21a}
&&\bar{S}_2(\omega)=-{V\over 2}\sum_{n,m}(2\pi)^2\int{d\nu\over 2\pi} dx_1 dx_2 {d\vp\over (2\pi)^2}
{d\vt\over (2\pi)^2} C_n^*(\omega)\times\nonumber\\
 &&C_m(\omega)\delta[t_y+eH(x_1+x_2)-kn]~ \delta(kn-km)\times \nonumber\\
&&{e^{it_x(x_1-x_2)}
\over [i(\nu+\omega)-\epsilon_\vp][-i\nu-(\epsilon_\vp-\vv_{\rm F}\cdot\vt)]}
\psi_n(x_1)\psi_m(x_2).
\end{eqnarray}
Here we have set $\vt\equiv\vq+\vp$ and
$\epsilon_\vq\simeq \epsilon_\vp-\vv_{\rm F}\cdot\vt$ with $\vv_{\rm F}$
the Fermi velocity.

Using $\int{d\vp\over (2\pi)^2}={\cal{N}}\int d\epsilon\int_0^{2\pi}
{d\phi\over 2\pi}$ (with ${\cal{N}}=m/2\pi$ as
 the 2D density of states) and the integral
($\Omega$ is the usual ultraviolet cutoff)

\begin{eqnarray} \label{eq:S22a}
&&\int_{-\infty}^\infty{d\nu\over 2\pi}\int_{-\Omega}^\Omega d\epsilon
{1\over [i(\nu+\omega)-\epsilon][-i\nu-(\epsilon-\vv_{\rm F}\cdot\vt)]}
\nonumber\\
&&=\ln\left[{4\Omega^2\over (\vv_{\rm F}\cdot\vt)^2+\omega^2}\right],
\end{eqnarray}
one obtains after some calculation

\begin{eqnarray} \label{eq:S23a}
\bar{S}_2(\omega)=K_2(\omega)\langle|\Delta_\omega(\vr)|^2\rangle,
\end{eqnarray}
where
$\langle|\Delta_\omega(\vr)|^2\rangle$ is defined in (\ref{eq:delta2}) and

\begin{eqnarray} \label{eq:S24a}
K_2(\omega)=-{g\over 4\pi eH}\int d\vt
e^{-{t^2\over2eH}}\ln\left[{4\Omega^2\over (\vv_{\rm F}\cdot\vt)^2+\omega^2}\right].
\end{eqnarray}
Here $g={\cal{N}}V/2$.
Expanding $K_2(\omega)=K_2(0)+A|\omega|+B\omega^2+...~$, we found the zero-frequency
term

\begin{eqnarray} \label{eq:K2a}
K_2(0)=-g\ln(2\Omega \tau_H)
\end{eqnarray}
with $\tau_H=\ell/v_{\rm F}$ and
$\ell=1/\sqrt{2eH}$,
and the first-order coefficient

\begin{eqnarray} \label{eq:Aa}
A=\left.{\partial K_2(\omega)\over \partial \omega}\right|_{\omega\rightarrow 0}
=\sqrt{\pi}g \tau_H.
\end{eqnarray}
The combination of (\ref{eq:K2a}) and (\ref{eq:Aa}) gives the
result in (\ref{eq:k2}).

\section{Derivation of $S_4$}
\label{appB}

The quartic action $S_4$ is expressed diagrammatically in Fig.~\ref{fig1}.
The similar diagram has been calculated by Maki and Tsuzuki \cite{maki65}
for 3D and finite-$T$. Here we need
$S_4$ in 2D and at $T=0$.
One can write the quartic term as
$S_4=\int d\omega d\omega_1 d\omega_2 \bar{S}_4(\omega,\omega_1,\omega_2)$,
where at zero frequencies ($\omega=\omega_1=\omega_2=0$)

\begin{eqnarray} \label{eq:S4a1}
&&\bar{S}_4(0)={V^2\over 4}
\int{d\nu\over 2\pi} d\vr_1 d\vr_2  d\vr_3 d\vr_4 G_{\nu}(\vr_1,\vr_2)
G_{-\nu}(\vr_3,\vr_2)\times\nonumber\\
&&G_{\nu}(\vr_3,\vr_4)
G_{-\nu}(\vr_1,\vr_4)\Delta^\dagger_0(\vr_1)\Delta_0(\vr_2)
\Delta^\dagger_0(\vr_3)\Delta_0(\vr_4).
\end{eqnarray}
In analogy to Appendix A, after a lengthy calculation we find

\begin{eqnarray} \label{eq:S4a2}
&&\bar{S}_4(0)={1\over 4}gV v_{\rm F}^{-2}{2\pi\over k}\sum_{n,m,p,q}
C_n^* C_m C_p^*C_q R(n,m,p,q),
\end{eqnarray}
where for brevity, $C_n\equiv C_n(0)$, and

\begin{eqnarray} \label{eq:S4a3}
&&R(n,m,p,q)=\delta(n-m+p-q)\int_0^{2\pi}{d\theta\over 2\pi}{1\over \cos^2\theta}
\times\nonumber\\
&&\int dx_1 dx_2 dx_3 dx_4
{\Theta(1,3;2,4)\over |x_1+x_3-x_2-x_4|}\times\nonumber\\
&&e^{-i\tan\theta eH T}
\psi_n(x_1)\psi_m(x_2)\psi_p(x_3)\psi_q(x_4).
\end{eqnarray}
Here $\theta$ is the azimuthal angle in 2D,

\begin{eqnarray} \label{eq:S4a4}
\Theta(1,3;2,4)\equiv &&
  1,~~ {\rm if}~ x_1,x_3>x_2,x_4 ~{\rm or}~x_1,x_3<x_2,x_4 \nonumber\\
&& 0,~~ {\rm otherwise}~,
\end{eqnarray}
and

\begin{eqnarray} \label{eq:S4a5}
&&T=-(x_1-{kn\over 2eH})^2+(x_2-{km\over 2eH})^2-(x_3-{kp\over 2eH})^2
\nonumber\\&&+(x_4-{kq\over 2eH})^2
+\left({k\over 2eH}\right)^2(n^2-m^2+p^2-q^2).
\end{eqnarray}
One can further solve

\begin{eqnarray} \label{eq:S4a6}
\bar{S}_4(0)=&&{1\over 4}gV v_{\rm F}^{-2}{2\pi\over k}\sqrt{\pi\over 4eH}
\sum_{n,m,p,q}\delta(n-m+p-q)\times\nonumber\\
&&e^{-Z}C_n^* C_m C_p^*C_q f,
\end{eqnarray}
where

\begin{eqnarray} \label{eq:S4a7}
Z&=&{k^2\over 4eH}\left[n^2+m^2+p^2+q^2-{1\over 4}\left
(n+m+p+q\right)^2\right]\nonumber\\
&=&{k^2\over 16eH}[(n-m)^2+(n-p)^2+(n-q)^2+\nonumber\\
&&~~~~~~~~~~~(m-p)^2+(m-q)^2+(p-q)^2]
\end{eqnarray}
and

\begin{eqnarray} \label{eq:S4a8}
f=&&\int_0^{2\pi}{d\theta\over 2\pi}{1\over \cos^2\theta}
\int_0^\infty t dt\int_0^1 du \int_0^u dv
e^{-eHt^2\Lambda}\nonumber\\
&&\cosh (\sqrt{eH}t\Sigma).
\end{eqnarray}
Here ($\alpha=\tan\theta$)

\begin{eqnarray} \label{eq:S4a9}
\Lambda={1\over 2}(1-i\alpha)(1-u)^2+{1\over 2}(1+i\alpha)v^2+
{1\over 4}(1+\alpha^2)
\end{eqnarray}
and

\begin{eqnarray} \label{eq:S4a10}
\Sigma=&&{k\over 2\sqrt{eH}}\times\nonumber\\
&&[(1-i\alpha)(1-u)(n-p)+(1+i\alpha)v(m-q)].
\end{eqnarray}

As pointed out by Maki and Tsuzuki, due to the factor of $e^{-Z}$
in (\ref{eq:S4a6}), there is a dominance of
$n=m=p=q$ when $k^2/4eH\gg 1$ ({\it i.e.}, in the weak-field limit).
In this limit $\Sigma\approx0$, and thus drops out of
$f$ in (\ref{eq:S4a8}). As a consequence, $f$ becomes independent of
$n,m,p,q$ and Eq.~(\ref{eq:S4a6}) is simplified to

\begin{eqnarray} \label{eq:S4a11}
\bar{S}_4(0)=&&{1\over 2}{g^2\over
{{\cal{N}}v_{\rm F}^2}}f\langle|\Delta_0(\vr)|^4\rangle,
\end{eqnarray}
where $\langle|\Delta_0(\vr)|^4\rangle$ is given in
(\ref{eq:delta4}).
To the leading order in field, we found

\begin{eqnarray} \label{eq:S4a12}
f=f(H)={4\pi\over eH}[\ln(1+\sqrt{2})]^2.
\end{eqnarray}
It is interesting to note that for 3D and finite-$T$,
the same term will be $f\propto -\ln (T/H) $,
and divergent at $T\rightarrow 0$
\cite{maki65}.


\begin{thebibliography}{10}
\bibitem{approx}
In general the result could be more complicated. We employ a similar
  approximation as that used in Ref.~\protect\cite{maki65} appropriate for low
  fields which leads to a complete independence of the coefficient on LLL (see
  also Appendix~\ref{appB}).
\end{thebibliography}

\end{document}